\input harvmac

\Title{
\vbox{\baselineskip12pt\hbox{hep-th/9702180}\hbox{HUPT-97/A010}
\hbox{UCB-PTH-97/11}\hbox{LBNL-40032}
}}
{\vbox{
\centerline{Geometry of $N=1$ Dualities in Four Dimensions} }}
\centerline{Hirosi Ooguri$^{1,2}$ and Cumrun Vafa$^3$}
\medskip
\centerline{$^1$ Department of Physics}
\centerline{University of California at Berkeley}
\centerline{Berkeley, CA 94720--7300}

\smallskip
\centerline{$^2$ Physics Division,
Mail Stop 50A--5101}
\centerline{
E. O. Lawrence Berkeley National Laboratory}
\centerline{Berkeley, CA 94720}
\centerline{\tt ooguri@physics.berkeley.edu}

\smallskip
\centerline{$^3$ Lyman Laboratory of Physics}
\centerline{Harvard University}
\centerline{Cambridge, MA 02138}
\centerline{\tt vafa@string.harvard.edu}
\bigskip
\centerline{\bf Abstract}
We discuss how $N=1$ dualities in four dimensions
are geometrically realized by wrapping D-branes about
3-cycles of Calabi-Yau threefolds.
In this setup the $N=1$ dualities for $SU$, $SO$
and $USp$ gauge groups with fundamental fields get
mapped to statements about the monodromy
and relations among 3-cycles of Calabi-Yau threefolds.  The
connection between the theory and its dual requires passing
through configurations which are T-dual to the well-known 
phenomenon of decay of BPS states in $N=2$ field theories in 
four dimensions. We compare our approach to recent works based
on configurations of D-branes in the presence of NS 5-branes
and give simple classical geometric derivations of 
various exotic dynamics involving D-branes ending on NS-branes.
%
\noindent

\Date{February 1997}
\newsec{Introduction}
Many of the field theory dualities have now been embedded into
string theory.  The basic idea is to construct a local
description of the field theory in a stringy setup.
This local description can involve either purely geometric
aspects of compactification
manifold \ref\kkv{
S. Katz, A. Klemm and C. Vafa, ``Geometric Engineering of Quantum Field
Theories,"
hep-th/9609239}, a local geometry together
with D-branes wrapped around cycles \ref\bjpsvz{
 M. Bershadsky, A. Johansen, T. Pantev, V. Sadov, C. Vafa,
 ``F-theory, Geometric Engineering and $N=1$ Dualities,"
hep-th/9612052\semi
C. Vafa and B. Zwiebach,``$N=1$ Dualities of 
$SO$ and $USp$ Gauge Theories and
T-Duality of String," hep-th/9701015}\
or D-branes in the presence of NS 5-branes
in a flat geometry \ref\hw{A. Hanany and E. Witten, ``Type IIB Superstrings,
BPS Monopoles,
and Three-Dimensional Gauge Dynamics," hep-th/9611230.},
\ref\dbhooy{J. de Boer,
K. Hori, H. Ooguri, Y. Oz
and Z. Yin, ``Mirror Symmetry in Three-Dimensional Gauge
Theories, $SL(2,Z)$ and D-Brane Moduli Space,"
hep-th/9612131, to appear in Nucl. Phys. B.},
\ref\egk{S. Elitzur, A. Giveon
and D. Kutasov, ``Branes and  $N=1$ Duality in String Theory,"
hep-th/970214.}.  In particular recently Elitzur,
Giveon and Kutasov \egk ,  following the approach of
Hanany and Witten in  constructing $N=4$ theories in $d=3$ \hw ,
found a rather simple description of how Seiberg's
$N=1$ duality in four dimensions arises.  They also suggested
that their approach is T-dual to that of \bjpsvz . However
their configuration of D-branes is more transparent
and it allowed them in particular
to see the appearance of the fundamental magnetic meson
field in a simple way.  In this paper we provide a local
geometric description with wrapped D-branes
in the spirit of \bjpsvz\ but in a somewhat simpler
way, for which one can also follow the D-brane configurations in detail
and see a particularly simple geometric realization
of Seiberg's duality.

We will also discuss how the approach of \egk\ is related to
the geometric description presented in this paper.
The advantage of \hw\ and \egk\ is that
the spacetime geometry is flat. On the other hand,
the dilaton field is not constant in the presence of
NS 5-branes, and in fact the string coupling
constant blows up at cores of the branes.
As noted in \hw , this makes it difficult to analyse exactly
what happens when D-branes end on NS 5-branes, which is a
typical situation in their cases.
For example, in their construction it was assumed
that an open string stretched between two 
D-branes ending on opposite sides of an NS 5-brane gives 
a matter multiplet. In such a situation, however, the
open string has to go through the strong coupling region
inside the core of the NS 5-brane and the derivation of this
statement would be beyond the reach of perturbative
string theory. In \hw , it was also suggested that
when D and NS-branes cross each other a third brane
should be created. This conjecture was motivated by 
comparison with field
theory results and on the conservation of the NS-NS charge. 
We will show that these and other exotic dynamics involving D-branes
ending on NS-branes have somewhat simpler counterparts in our
construction and can be explained in purely
classical geometric terms.
Moreover the geometrical approach we follow easily
extends to $SO$ and $USp$ gauge theories similar to
\bjpsvz .

\newsec{Geometrical Setup}

Compactification of type II strings on Calabi-Yau
threefolds leads to $N=2$ theories in $d=4$. We will be
interested in a local model for such a compactification
which corresponds to a non-compact Calabi-Yau threefold.
A canonical class of BPS states corresponds to
Dirichlet $p$-branes wrapped around $p$-cycles of the
Calabi-Yau.  They preserve $1/2$ of the supersymmetry,
i.e. on their worldline we obtain the reduction of an $N=1$ system
from four dimensions to $(0+1)$-dimension.  If we consider
the spatial directions to be a $T^3$ and T-dualize
(exchanging IIA and IIB strings)
we end up with $(p+3)$-branes partially wrapped around
cycles of Calabi-Yau threefold, and at the same time filling
the spacetime.  The theory living on the (3+1)-part of the
spacetime worldvolume
of the $(p+3)$-brane is an $N=1$ theory in $d=4$.  In this
way we have mapped BPS states of an $N=2$ string theory to
 $N=1$ field theories in four dimensions\foot{More generally
this connection
may provide an interesting link between black-hole
dynamics in various dimensions and the T-dual field theories.}.

We would now like to explore some aspects of the resulting
field theory in connection with the D-brane configurations.
Let us consider type IIB on Calabi-Yau threefold, and
consider some number of D3-branes wrapped around a set of
three cycles $C_i$ of Calabi-Yau
threefold.  Let $\omega$ denote the holomorphic
3-form of the Calabi-Yau.  For a set of 3-cycles
such a configuration can correspond to a BPS state
only if \ref\bpsch{
 A. Ceresole, R. D'Auria, S. Ferrara and A. Van Proeyen,
``Duality Transformations in Supersymmetric Yang-Mills Theories Coupled to
  Supergravity,"
Nucl. Phys. B444 (1995) 92, hep-th/9502072}
\eqn\bpsc{|\sum_i \int_{C_i}\omega| =\sum_i |\int_{C_i}\omega|}
i.e. the vectors $\int_{C_i}\omega$ in the complex plane are all
parallel.
If we consider T-dualizing the 3-space, we end up with
type IIA theory with D6-branes wrapping around 3-cycles $C_i$
of Calabi-Yau
and filling the spacetime.  Again the condition
\bpsc\ is
the condition corresponding to having an $N=1$ supersymmetric
field theory in $d=4$.
There are two natural classes of 3-cycles that appear in
Calabi-Yau threefolds: (A) $S^2\times S^1$ and
(B) $S^3$.
Moreover in a neighborhood of these cycles the Calabi-Yau
threefold can be approximated by the cotangent space
$T^*(S^2\times S^1)=T^* S^2\times S^1\times R $ and $T^*S^3$.
In case (A), the situation is locally the same as
D-branes wrapped around $S^2\times S^1$ in $K3\times T^2$ compactification,
where we view $T^*S^2$ as part of $K3$ and $T^*S^1$ as
part of $T^2$.  In this case the field theory in $d=4$ will thus
have $N=2$ instead of $N=1$.  In fact if we consider
$N$ D-branes wrapped around such a cycle we end up getting
an $N=2$ system with $U(N)$ gauge symmetry and no matter
\ref\bsv{M. Bershadsky, V. Sadov and C. Vafa, ``D-Branes and Topological
Field Theories,'' Nucl. Phys. B463 (1996) 420, hep-th/9511222.}.
In $N=1$ terminology this is the same as an $U(N)$ gauge
system with an adjoint chiral multiplet.
If we wrap $N$ D-branes around cycles of type (B) we
end up with a pure $N=1$ gauge system with gauge group $U(N)$.
Note that in either case the bare gauge coupling constant
is related to the volume of the three-cycle $C_i$ by
$${1\over g^2}\sim V_{C_i}$$
which follows from the contribution to gauge coupling
constant from 7 to 4 dimensions upon compactification on
the 3-cycle. We have to note, however, this formula
can get strong quantum corrections when $V_{C_i}$ is small.

If a pair of wrapped cycles $C_i, C_j$ intersect one another, the
corresponding wrapped D-branes will be intersecting, in which case
we can obtain extra massless matter from open strings
ending on the pair of D-branes.  For the intersection
to be supersymmetric (and in particular
to be compatible with \bpsc ) we need that the number of
local Dirichlet versus Neumann boundary conditions
for the open string sector to be 0 mod 4, which
in the present context means that the cycles $C_i$ and $C_j$
intersect on a circle.  If we have $N_i$ D-branes wrapped
around $C_i$ and $N_j$ D-branes wrapped around $C_j$
in such a situation
the open string sector will give us a chiral matter
of the type $(N_i,N_j)$ (i.e. one $N=1$ chiral
multiplet in the fundamental representation 
in $U(N_i)$ times the conjugate representation
in $U(N_j)$). We will refer to it as bifundamental.
If $C_i$ is of $A$ type, we in addition will have
a superpotential interaction of the form
 $qM\tilde q$ where $(q,\tilde q)$ correspond to
chiral matter matter and
$M$ is the adjoint matter of $U(N_i)$ coming from the D-branes wrapped
around $C_i$.  This follows from the fact that the theory as seen
from the D-branes wrapped around $C_i$ has an $N=2 $ supersymmetry.

We shall be interested in changing the complex moduli
of Calabi-Yau threefold and following what happens to the
wrapped cycles and discuss the corresponding field theory
interpretation.  In particular we shall consider
a situation where cycles of both (A) and (B) type
 appear.  Our local description of
Calabi-Yau is that given in \ref\bsvds{M. Bershadsky,
V. Sadov and C. Vafa, ``D-Strings
on D-Manifolds,'' Nucl. Phys. B463 (1996) 398, hep-th/9510225.}\
which we now
review.  Consider local coordinates of the Calabi-Yau
3-fold given by
$(x,y;x',y';z)$ subject to two relations:
$$x^2+y^2=\prod_i (z-a_i)$$
$$x'^2+y'^2=\prod_j (z-b_j)$$
This geometry can be viewed more abstractly as
a $C^*\times C^*$ bundle over the $z$-plane,
viewing $(x,y)$ and $(x',y')$ as coordinates
of the $C^*$'s; we can in turn view each $C^*$ as
approximating the structure of an elliptic curve near
its degeneration.  Note that the total space is non-singular
if all $a_i$ and $b_j$ are distinct; the local degeneration
of the fibers is an artifact of how we are slicing the total
space.  The space becomes singular if any pair of the $a_i$'s or
$b_j$'s coincide where we get some vanishing cycles.
Let us see how these cycles arise:
To any pair of $a_i$'s (and similarly $b_j$'s) we can associate
a 3-cycle of type (A) and to any $(a_i,b_j)$ pair
we can associate a 3-cycle of type (B).  To see this
note that for a fixed $z$, away from $a_i$ and $b_i$
there is a non-trivial $S^1$ in each of the $C^*$'s.
For example the equation $x^2+y^2={\rm const.}$
defines a circle (note that if the constant is
a positive real number this is realized by taking
$x$ and $y$ real. Otherwise by an overall change
of phase of $x$ and $y$ the situation reduces to the above
case).   Note that if we are at $z=a_i$
(or $z=b_j$) the corresponding
circle vanishes.
 We consider 3-cycles which are a product of $S^1\times
S^1$ cycles over each point on the $z$-plane, together with segments in the
$z$-plane ending on the $a_i$ or $b_j$. If we go between two $a_i$'s
without going through $b_j$ the corresponding three cycles
sweep out an $S^2\times S^1$ (and similarly if we go
between any two $b_j$'s).  However if we go between $a_i$
and $b_j$ the three cycle we obtain is an $S^3$. To see this
note that by continuous deformation the
situation is the same as the case where $a_i$ is
close to $b_j$ in which case locally the situation is the
same as
$$x^2+y^2\sim z-a \qquad x'^2+y'^2\sim b-z$$
which implies that
$$x^2+y^2+x'^2+y'^2\sim b-a$$
which clearly describes an $S^3$ (with no loss of generality
take $b$ and $a$ to be real and take $(x,y)$ and $(x',y')$
also to be real).

Connecting the pairs of $a_i$ and $b_j$ by paths
in the $z$-plane we can associate 3-cycles to each path.
Let us denote the 3-cycle connecting $a_i$ to $b_j$ by $[a_i,b_j]$.
An important aspect of the above geometry that we shall use
later is that the three cycle $[a_i,b_j]+[b_j,a_k]=[a_i,a_k]$.
This in particular means that the sum of two 3-cycles of type
(B) cycles can be a cycle of type (A).  This is actually
T-dual to the statement that a vector multiplet can decay
to hypermultiplets in the $N=2$ situation, as is well known
in field theory
\ref\swi{N. Seiberg and E. Witten,``Electric-Magnetic Duality,
Monopole Condensation and Confinement in $N=2$ Supersymmetric
Yang-Mills Theory," Nucl. Phys. B426 (1994) 19, hep-th/9407087}\
and its stringy realization \ref\ler{A. Klemm, W. Lerche,
P. Mayr, C. Vafa and N. Warner, ``Selfdual Strings and $N=2$
Supersymmetric Field Theory,'' Nucl. Phys. B477 (1996) 746,
hep-th/9604034.}.

Finally we wish to compute the integral
of the holomorphic 3-form $\omega$ over the $S^1\times S^1$
fiber over each point on $z$-plane and obtain a 1-form.
This is similar to the situation studied in  \ler\ and one finds (by a suitable
choice for $\omega$) we have
$$\int_{S^1\times S^1} \omega=dz$$
This in particular means that if we wish to have the condition \bpsc\
satisfied, the beginning and end point cycles that we end upon
must be in the same direction, i.e. $a_i-a_j,a_i-b_j, b_i-b_j$
must all be parallel if they correspond to the end points of the
cycles which have D-branes wrapped around.  Moreover to minimize
the volume of the cycle we must take the image of $D$-branes
on the $z$-plane to be straight lines.

\newsec{Geometric realization of $N=1$ dualities}
Consider the geometric setup described in the previous
section.  Suppose we have two points $a_1$ and $a_2$ along the
real part of $z$-plane where the first $C^*$ degenerates
and one point $b$, again on the real axis {\it between}
$a_1,a_2$ where the second $C^*$ degenerates.
In particular along the real axis we have three ordered
special points $a_1,b,a_2$.  Suppose we wrap $N_c$ D-branes
around the $S^3$ cycle $[a_1,b]$ and $N_f$ D-branes around the $S^3$ cycle
$[b,a_2]$.  Note that $[a_1,b]$ and $[b,a_2]$ meet along the circle
on the first $C^*$ at $z=b$.  Thus from the considerations
of the previous section it follows that the field theory we end
up with is given by an $N=1$ gauge theory
$U(N_c)\times U(N_f)$
with chiral matter in $(N_c,  N_f)\oplus {\rm c.c.}$ representation.
We will assume $N_f\geq N_c$.  Note that the above system
is the same as $N=1$ supersymmetric QCD where we have also
gauged the flavor group.

We now wish to change the moduli of Calabi-Yau and come
to a configuration where the degeneration points are still along the
real $z$-axis but the orders have been changed
from $(a_1,b,a_2)$ to $(b,a_1,a_2)$.  To do this we first
push the point $b$ up along the imaginary direction.
It is now energetically preferable for the D-branes to reconnect
so that $N_c$ of them go directly between $(a_1,a_2)$, by combining
$N_c$ pairs of $S^3$ cycles and converting them to
 $N_c$ cycles of $S^2\times S^1$ type,
and
$(N_f-N_c)$ of them go between $(b,a_2)$.  Now we push $b$
along the negative real axis so that it has passed the $x$-coordinate
of $a_1$ and then we bring it down to the real axis. At this
point the $(N_f-N_c)$ D-branes which were going between $(b,a_2)$
will decompose to $(N_f-N_c)$ branes between $(b,a_1)$ and $
(N_f-N_c)$ D-branes between $(a_1,a_2)$.  The $N_c$
branes we previously had along $(a_1,a_2)$ will recombine
with the new $(N_f-N_c)$ D-branes giving a total of
$N_f$ D-branes along $(a_1,a_2)$ cycle.  Thus the final configuration
we end up with is a configuration of points
ordered as $(b,a_1,a_2)$ with $(N_f-N_c)$ D-branes wrapped
around $(b,a_1)$ and $N_f$ D-branes wrapped around $(a_1,a_2)$
cycle. The field theory we end up with is again easy
to read off from the discussion of the previous section, namely
$$U(N_f-N_c)\times U(N_f)$$
with matter $q,{\tilde q}$ in $(N_f-N_c,N_f)\oplus {\rm c.c.}$ representation
and in addition, since the $(a_1,a_2)$ system is an $N=2$
system we have an extra adjoint field $M$ which interacts
with the above quarks in the usual form dictated by
$N=2$ supersymmetry, i.e. with a superpotential $qM{\tilde q}$.
We have thus transformed the supersymmetric QCD with gauged
flavor group to the Seiberg dual \ref\sei{N. Seiberg,
``Electric-Magnetic Duality in Supersymmetric
Nonabelian Gauge Theories,'' Nucl. Phys. B435 (1995)
129, hep-th/9411149.}\
where the flavor group continues to be gauged
(note that on the magnetic dual side the flavor
gauge system is an $N=2$ system, as we have above).

One may ask what is the field theoretic meaning
of the above operation.  This is simply turning
on the FI D-term for the $U(1)$ (common to the
flavor and the color group).  This breaks
supersymmetry completely which is reflected by the
fact that the intermediate D-brane configurations we were
considering were not parallel.  One may wonder
if we can pass only through supersymmetric
preserving configurations.  This can be done in two
ways.  One way is to just pass the point $b$ over
$a_1$ along the real axis, in which case the
conservation of D-brane charge will tell us
how many D-branes we will have wrapped
around each cycle after we pass through
the singular configuration.  Another, and perhaps
a more satisfactory description is to take the point
$a_2\rightarrow \infty$, in which case the flavor
gauge group coupling goes to zero and thus becomes just
a global symmetry group throughout the above process.
  In this case the D-brane
configurations will not break supersymmetry, because
in this limit the
lines on the $z$-plane are parallel, in accord with the fact that
in this case the $U(1)$ FI D-term does not break 
supersymmetry\foot{There is yet another way to turn on the FI D-term without
breaking supersymmetry. To do this, we add one more point
$a_3$ on the real axis to the right of $a_2$ and allow
the first $C^*$ to degenerate there also. We then wrap
additional $N_c$ D-branes on the $S^2 \times S^1$ cycle
$[a_2, a_3]$. In this case, we can lift the $S^3$ cycle $[b,a_2]$
off the real axis together with $(N_f - N_c)$ D-branes on it
while keeping $(a_2 - b)$ parallel to the real axis.
Note that we can now do this without sending
$a_2$ and $a_3$ to $\infty$. The field theory counterpart
of this construction is to consider a theory with
$U(N_c) \times U(N_f) \times U(N_c)$ gauge group. It is easy to
see that this contains a FI-parameter corresponding to
the lifting of the D-branes which does not break supersymmetry.
We can push this line of argument further and consider
$N_f$ ordered points $a_2, ... , a_{N_f +1}$ to the right of $b$,
with $(N_f - n)$ D-brane wrapping on the cycle
$[a_{n+1}, a_{n+2}]$ ($n=1,...,N_f-1$). By taking all
$a_{n+1} \rightarrow \infty$, we recover the $U(N_c)$
gauge theory with $N_f$ quarks, but this construction
allows us to give a different mass parameter to
each quark.}.

Note that the same manipulations as above would have also worked
if we had considered the $N=2$ configuration with $(a_1,a_2,a_3)$
in which case, in the limit we freeze the flavor group we would
have connected the $N=2$ system
$$U(N_c)\rightarrow U(N_f-N_c)$$
each with $N_f$ fundamental hypermultiplets.

\newsec{Generalizations to $SO$ and $USp$ theories}
In this section we will generalize the above construction
to the case of $SO$ and $USp$ gauge theories, very much in the
spirit of the second reference in \bjpsvz , and obtain
the $N=1$ dual pairs proposed in \sei , \ref\insei{K. Intriligator
and N. Seiberg, ``Duality, Monopoles, Dyons, Confinement and
Oblique Confinement in Supersymmetric $SO(N_c)$ Gauge Theories,"
Nucl. Phys. B444 (1995) 125, hep-th/9503179.},
\ref\inpo{K. Intriligator
and P. Pouliot,``Exact Superpotentials, Quantum Vacua and
Duality in Supersymmetric $SP(N_c)$ Gauge Theories,"
Phys. Lett. B353 (1995) 471, hep-th/9505006.}.

We start with the same setup as in the $SU$ case and consider
the double fibration
$$x^2+y^2=-(z-a)(z-a')$$
$$x'^2+y'^2=-z$$
where we take $a,a'$ as real numbers with $a<0$ and $a'>0$.
We thus have two $S^3$ cycles $[a,0]$ and $[0,a']$.
Note that the $S^3$ associated with $[a,0]$
is realized by considering
real values for $x,y,x',y',z$ (because for $a<z<0$
both $x^2+y^2$ and $x'^2+y'^2$ are positive), however the
$S^3$ associated with $[0,a']$ is realized by purely
imaginary values for $x,y,x',y'$ but real value for $z$.

We wrap $N_c$ D6-branes around $[a,0]$ and $N_f$
D6-branes around $[0,a']$. Now we orientifold
the above configuration by combining the operation
$$(x,y,x',y',z)\rightarrow (x^*,y^*,x'^*,y'^*,z^*)$$
with exchange of left- and right-movers on the worldsheet.
This is a symmetry of the above equation for $a$
and $a'$ real.  Note that the fixed point space, i.e.
the orientifold 6-space, is precisely
the first $S^3$ associated with $[a,0]$
times the uncompactified spacetime.
Under this orientifolding the groups we started with
$U(N_c)\times U(N_f)$ now change either to
$SO(N_c)\times USp(N_f)$ or $USp(N_c)\times SO(N_f)$,
depending on the choice of the sign for the cross cap diagrams,
with matter in bifundamentals as before.  In the terminology
of type I' theory it is natural to count the D-branes after
orientifolding as ${1 \over 2} N_c$ and ${1 \over 2} N_f$ 
D6-branes respectively.
Note that the reduction of the gauge factor associated with the 
${1 \over 2} N_f$ D6-branes
arise because of the action of the orientifolding on the D6-branes
wrapped around $[0,a']$ cycle; even though this cycle is not fixed
under this transformation pointwise it is still mapped to itself.

Let us note that for the net D6-brane charge on the $[a,0]$ cycle,
 in addition to the contribution from the physical
D6 branes, there is a contribution from the orientifold plane.  Since we
have an orientifold 6-space this contribution is $ \mp 16/2^3=\mp 2$ (i.e.
down from the case of orientifold 9-space by a factor of $2^3$
arising from T-duality 3 times each of which splits it to two copies).
The $\mp$ sign refers to the $SO(N_c)$ versus $USp(N_c)$ cases respectively.
Thus the net $D6$-brane charge of the $[a,0]$ cycle is ${1 \over 2}
N_c \mp 2$.The D6-brane charge of the $[0,a']$ cycle is ${1 \over 2}
N_f$ as there is no additional orientifold contribution to it.

Now we try to repeat the same process as in the $U(N_c)$ case.
The main difference here is that we cannot lift the degeneration
points off the real axis, as that is not consistent with the
orientifolding operation.  This is in accord
with the field theory description where in this case
we do not have the freedom to turn on a FI D-term.
Instead we take $a$ along the real
axis from negative to positive values.
After $a>0$ the $S^3$ represented by the $[a,0]$ for $a<0$
now becomes
an $S^3$  representing $[0,a]$
with purely imaginary values for $x,y,x',y'$.  In particular
the orientifolding operation has no fixed points anymore
but the gauge group still continues to be $SO$ or $USp$ due
to the action of the orientifold group on it.
To find out how many D-branes  we have wrapped around
$[0,a]$ and $[a,a']$ we simply use charge conservation;
this is an assumption which strictly speaking we cannot
prove because we have passed through a strong coupling region,
however the experience of the $U(N_c)$ case shows that it is reasonable.
Taking into account the orientations of the D-branes we now should
have ${1 \over 2} N_f -({1 \over 2} N_c \mp 2)$ D6-brane charge
on $[0,a]$ and ${1 \over 2} N_f$ D6-branes on $[a,a']$.
Noting that for $a>0$ there is no orientifold plane
all these charges should be accounted for by physical
D-branes, and thus we obtain the dual groups
$SO(N_f-N_c+4)\times USp(N_f)$ or $USp(N_f-N_c-4)\times SO(N_f)$
again with bifundamentals.  Moreover, just as in the $U(N_c)$
case we will again obtain an $N=2$ system in the flavor
group, which implies that we have the fundamental magnetic meson with
the right interactions.

\newsec{Does this prove $N=1$ duality?}

In the $N=1$ context the above process connects
the electric gauge system with its magnetic dual.
In what sense does this prove they are equivalent?\foot{We have
benefited from discussions with N. Seiberg in preparation
of this section.}
Note for example, in the context of heterotic string compactifications
on $T^2$, the fact that we can continuously connect an $SU(3)$
gauge system with an $SU(2)\times SU(2)$ gauge system does not imply their
equivalence.  In fact as discussed above in the context of $N=2$
systems we have connected a $U(N_c)$ system with an $U(N_f-N_c)$ system
which clearly are inequivalent (for $N_f\not=2 N_c$) as they
even have different dimensions for their Coulomb branch.

One hint of how one may try to understand in which cases we should expect
an equivalence is that if in the process of exchange
we had not pushed the middle point off the real axis and
just gone along the negative real axis
 to the point where it would meet the first
degeneration point, the original theory and the dual theory would meet
and become the same theory at that point. This is the point where the gauge
coupling
constant in both theories are going to infinity.
  Now if we take into account the quantum corrections,
and assuming both the original and the dual theories are asymptotically
free (in the non-trivial
$SU$ part of the gauge group), taking the infrared limit on both theories will
push
both to the strong coupling regime where we can see how they
can become equivalent.  Of course this is a region where
we should expect strong quantum corrections to the classical
D-brane picture; however the above heuristic argument seems
to at least give a conservative rationale to indicate in
which cases the above interpolation between
theories may imply infrared equivalence.  Note that in the $N=2$ case
either the original or its dual are not in the asymptotically
free regime, except for $N_f=2N_c$ ( in the $SU$ case) where the above
equivalence is the conjectured S-duality of $N=2$
systems, the above connection
would not necessarily imply their equivalence in the infrared,
thus avoiding a contradiction.
However in the $N=1$ with $SU(N_c)$ gauge group
 for ${3 \over 2} N_c <N_f<3N_c$ since
both the original and the dual theory are asymptotically
free the above interpolation between theories suggests
their infrared equivalence.  It does
not seem clear to us why in the regime $N_f< {3 \over 2}N_c$ or
$N_f>3N_c$ where either the magnetic or the electric system
is not in the asymptotically free region the
above interpolation shows their infrared equivalence.

\newsec{Relation to other approaches}

As noted before our approach is similar in spirit
to that of \bjpsvz .  In particular in the $N=2$ situation
it is identical to it by T-duality:  If we consider type IIB
compactified on $K3$ and consider $N_c$ D7-branes
wrapped around $K3$ and $N_f$ D3-branes on it,
the process considered in \bjpsvz\ consists of
taking the volume of $K3$ to be small and using
T-duality, exchanging 0- and 4-cycles on $K3$, to obtain the induced
D-brane charges\foot{These aspects are
studied further in \ref\horioz{K. Hori and Y. Oz, ``$F$-Theory,
T-duality on $K3$ Surfaces and $N=2$ Supersymmetric Gauge
Theories in Four Dimensions,'' hep-th/9702173.}.}.  
This however can be simplified
by noting that the $SO(20,4; {\bf Z})$ T-duality
on $K3$ \ref\asmo{P.S. Aspinwall and D.R. Morrison,
``String Theory on K3 Surfaces," hep-th/9404151}\
maps the above process into the classical
monodromy of $2$-cycles\foot{To be more specific,
the T-duality which exchanges the 0 and the 4-cycles
is conjugate, by the mirror symmetry, to the
classical monodromy of $K3$.}. This also maps 3-brane
and 7-brane configuration to 5-brane configurations
wrapped around two-cycles of $K3$. By T-duality
around one extra circle, this is exactly the configuration
we have considered in the previous section in the $N=2$
case.

However our construction of the $N=1$ case
seems more difficult to relate to \bjpsvz , and
in particular for the case of $SU$ gauge groups it is more closely
related to the recent construction of Elitzur, Giveon and Kutasov
\egk .

\subsec{From Calabi-Yau to multiple brane configuration}

The connection of our approach to that of \egk\
becomes apparent when we note
that the $A$-type singularity on $K3$,
\eqn\nsbrane{ x^2 + y^2 = \prod_i (z-a_i) }
is related, by T-duality, to a configuration of parallel NS 5-branes
\ref\ov{H. Ooguri and C. Vafa, ``Two-Dimensional Black Hole and
Singularities of CY Manifolds," Nucl. Phys. B463 (1996) 55,
hep-th/9511164.}.
This can be shown by performing the T-duality on the elliptic fiber,
along the natural $S^1$ action on $C^*$.
In the original geometry \nsbrane , the elliptic fiber undergoes a
monodromy transformation
$\tau \rightarrow \tau +1$ around each point $z=a_i$. After the
T-duality, exchanging type IIA and type IIB,
this becomes a unit integral shift in
the NS-NS $B$-field on the fiber. Therefore the integral
of $H = dB$ on a small circle around $z=a_i$ times the fiber
gives $1$, namely the region near $a_i$ carries the minimum
unit of the NS-NS charge.
Note that the dilaton gets turned on in this process
since the radius of $S^1$ on the fiber
depends on the position $z$ on the base.
This shows that the T-duality replaces
the degeneration of the fiber at each $a_i$ by one NS 5-brane.

We can also perform the T-duality on each $C^*$ of
the double elliptic fibration (which
now takes type IIA or IIB back to itself),
\eqn\double{\eqalign{
  x^2 + y^2 & =  \prod_i (z-a_i) \cr
  x'^2 + y'^2 & =  \prod_i (z-b_i) ,\cr}}
giving rise to two types of NS 5-branes, oriented differently.
Let us choose coordinates so that NS 5-branes of
the first type are parallel to the $x^0, ..., x^3, x^4, x^5$ plane,
and NS 5-branes of the second type are stretched in
the $x^0,..., x^3, x^8, x^9$ directions.
Since $x^6, x^7$ are common transverse directions
to both types of NS 5-branes, we may regard $(x^6, x^7)$ as real
and imaginary parts of $z$ in \double. Therefore
$x^6 + i x^7 = b_i$ for a location of an NS 5-brane of the first type and
$x^6 + i x^7 = a_i$ for the second type.
The type II string on this geometry would give an $N=2$ theory
in four dimensions in the $x^0,..., x^3$ directions. Following
\egk , we refer NS 5-branes of the first and second types as NS
and NS'-branes respectively.

Let us consider
D6-branes wrapping on the $S^2 \times S^1$ cycles
$[a_i,a_j]$, $[b_i,b_j]$ or on the $S^3$ cycles $[a_i,b_j]$.
Since these D6-branes locally look like $S^1 \times S^1$
on the fiber times line segments on the base $z$-plane, the T-duality
on the fiber squeezes the $S^1 \times S^1$ directions on the
branes and leaves them stretched on the line segments on the base.
Thus the D6-branes turn into D4-branes connecting the NS 5-branes.
Earlier in this paper, we found $a_i - a_j, b_i- b_j, a_i - b_j$
must all be parallel when there are D6-branes wrapping on the
corresponding cycles. From the T-dual picture, the reason for this
is that all the D4-branes have to be parallel in order to preserve the
$N=1$ supersymmetry. We choose coordinates so that this direction
is parallel to the $x^6$ axis, i.e. $a_i-a_j, b_i-b_j, a_i-b_j$
are constrained to be real.

The geometric realization of the $N=1$ $U(N_c) \times U(N_f)$ gauge
theory with chiral matter in $(N_c , N_f) \oplus {\rm c.c.}$
in the previous section is then mapped to the configuration
of D-branes in the presence of the NS 5-branes.
By reading from the right to left along the $x^6$ axis,
an NS'-brane located at $a_2$ on the base
is connected by $N_f$ D4-branes to an NS-brane at $b$
which is then connected by $N_c$ D4-branes to another NS'-brane at $a_1$.
One recognizes that this configuration is similar to that
of \egk\ except that, in their case, the role of the right-most
NS'-brane is played by $N_f$ D6-brane stretched in the $x^0, ... ,
 x^3, x^7, x^8, x^9$ directions.

Let us compare the two approaches. In \egk, one has to make
an assumption about a configuration which involves D4-branes
ending on an NS-brane. For example, it is assumed
that an open string stretched between two D4-branes attached on
opposite sides ($x^6 < b$ and $b < x^6$) of the NS-brane
gives the matter in $(N_c, N_f)$. However, as noted in
\hw , such an open string has to go through the core of 
the NS-brane where the string coupling constant blows up, and 
it is difficult to see what exactly is happening there. 
This issue is avoided in our construction since the dilaton 
is constant. Moreover the total space of the elliptic
fibration is non-singular even at $z=b$. 

There are other interesting dynamical effects
associated to the presence of NS 5-branes.
It was suggested by Hanany and Witten \hw\
that, when the D6-brane crosses the NS-brane by cutting
through it, an extra D4-brane should be created
between the D6 and NS-branes. This conjecture was
motivated by the consistency with field theory results and
the conservation of the NS-NS charge. Similarly
they argued that, if there are more than one D4-branes stretching between
the D6 and NS-branes, the resulting configuration
(called the s-configuration in \hw) should not have
a supersymmetric ground state. We will show below that the 
corresponding statements in our setup can be explained by geometric
terms.

\subsec{Geometric derivation of the Hanany-Witten effect}

What happens when a D6-brane stretched in the $x^0,...,
 x^3, x^7, x^8, x^9$ directions crosses an NS-brane stretched
in the $x^0,..., x^3, x^4, x^5$ directions? According to Hanany
and Witten, there must appear a D4-brane parallel to
the $x^0,..., x^3, x^6$ plane and connecting the
D6 and NS-branes\foot{To simplify our notations,
we ignore the common $x^0,..., x^3$ directions in the following
discussions.}.

To understand its geometric meaning,
let us perform the T-duality back to the double elliptic
fibration of the Calabi-Yau manifold.
Let us call the homology 1-cycles on the first elliptic
fiber $\alpha_1$ and $\beta_1$, and the cycles
on the second fiber $\alpha_2$ and $\beta_2$.
We choose the basis of the cycles so that, after the T-duality,
the $\alpha_2$-cycle vanishes at $b = x^6 + i x^7$ where the NS-brane
was located. The D6-brane in question is
localized in the $x^4, x^5$ direction, i.e. on the first elliptic
fiber, and is wrapping on the entire second fiber.
It is also stretched along the $x^7$ direction.
Therefore after the T-duality on $\alpha_1$ and
$\alpha_2$, this D6-brane transforms itself into another D6-brane
now wrapping on the $\alpha_1$ and $\beta_2$ cycles
and stretched in the $x^7$ direction.

Since the geometry is asymptotically locally Euclidean,
we can impose boundary conditions for large $x^7$ so that
the D6-brane configuration looks asymptotically like $\alpha_1 \times
\beta_2 \times ({\rm the}~ x^7 ~{\rm direction})$.
Let us move the D6-brane along the $x^6$ axis
toward $z=b$ and see what happens. We should note
that the local degeneration of the fiber at $z=b$ is
an artifact of how we are slicing the total space,
and there is no geometric singularity at $z=b$. 
Therefore we should
be able to describe the passing of the D6-brane through
$z=b$ by purely geometric language and the change
of its shape should be smooth. Now let us  push $x^6$ 
to the other side of $b$
while keeping these boundary conditions at 
$x^7 \rightarrow \pm \infty$. Because of 
the monodromy $\beta_2 \rightarrow \beta_2 + \alpha_2$ around
$z=b$, with an appropriate marking of the cycles on the fiber,
a cross section of the D6-brane configuration for
fixed $x^7$ right above $x^7=0$ is now
$[\alpha_1 \times (\beta_2 + \alpha_2)]$ while a cross
section right below $x^7=0$ remains $[\alpha_1 \times \beta_2]$.
They do not match at $x^7 = 0$. The only thing that
can happen is that the $x^7>0$ and $x^7 < 0$ portions
of the D6-brane combine to create another D6-brane wrapping on
$[\alpha_1 \times \alpha_2]$ at $x^7 = 0$ through
a {\it pants-diagram}.  The new D6-brane then can
go from $x^6$ to $b$ where $\alpha_2$ is annihilated.
This new portion of the D6-brane has topology of
a solid torus whose boundary is $[\alpha_1 \times \alpha_2]$
at $x^6$ and the $\alpha_2$-cycle is contractible
inside of the solid torus. The boundary of this solid torus
fills the mismatch of the $x^7>0$ and $x^7 <0$ portions of the
D6-brane, and the resulting configuration is
supersymmetric and of minimal volume with respect to
the boundary conditions at $x^7 \rightarrow \pm \infty$
given in the above.
After performing the T-duality on $\alpha_1$ and $\alpha_2$,
a portion of the D6-brane wrapping on
the solid torus turns into a D4-brane on the
line segment $[x^6, b]$. We see that
this is exactly the configuration conjectured
in \hw, i.e. the D6-brane is now connected by a D4-brane
to the NS-brane.

{}From this discussion, it is also clear why the configuration
with more than one D4-branes going between the D6 and NS-branes
(called the s-configuration in \hw) is not supersymmetric.
The corresponding configuration in our setup would involve
two portions of D6-branes whose cross sections
for fixed $x^7$ are $[\alpha_1 \times \beta_2]$ at $x^7<0$ and
$[\alpha_1 \times (\beta_2 + n \alpha_2)]$ at $x^7>0$
with $n>1$. To tie them together at $x^7=0$, we need
a solid torus whose boundary is $[\alpha_1 \times n \alpha_2]$.
However we cannot set it in between $x^6$ and $b$ without
creating a curvature singularity at $z=b$.

Thus the entire construction of \hw\ and \egk\
is mapped into geometrical language we have been considering.

\newsec{Comment on the instanton moduli space on the ALE space}

We would like to
comment on Kronheimer-Nakajima's construction \ref\kn{
P.B. Kronheimer and H. Nakajima, ``Yang-Mills Instantons
on ALE Gravitational Instantons,'' Math. Ann. 288 (1990) 263.}
of the instanton
moduli space on the ALE space since
a geometric construction similar to those discussed in
the above gives a natural D-brane
interpretation of their result\foot{We would like
to thank M. Douglas and N. Seiberg for discussion on
this subject.}. According to Kronheimer and Nakajima,
the moduli space ${\cal M}_k(V)$ of instantons of degree
$k$ on a vector bundle\foot{${\cal R}_i$ ($i=1,...,n$) 
are particular line bundles
over an ALE space of the $A_{n-1}$ type associated to
the different representation of ${\bf Z}_n$} 
$V = \oplus {\cal R}_i^{\otimes v_i}$
with gauge group $U(V)$ is
the largest Higgs branch of the $N=4$ gauge theory
in three dimensions with gauge group $\prod_{i=1}^n U(k)_i$
with $v_i$ hypermultiplets in $k$ of
$U(k)_i$ and one bifundamental with respect to
of $U(k)_i \otimes U(k)_{i+1}$.

This $N=4$ gauge theory can be obtained by
compactifying the type II string on $K3 \times T^3$ and
wrapping D4 branes on 2-cycles on $K3$ localized at points
on $T^3$. The relevant local model is again the elliptic
fibration over the $z$-plane, but in order to reproduce the
field content we compactify the real part of
$z$ on $S^1$ and pick $n$-points $a_1, ... , a_n$ on $S^1$
where the fiber degenerates. There are $n$ $S^2$-cycles
on this space, $[a_1, a_2]$, $[a_2, a_3]$, ... , $[a_n, a_1]$,
and we wrap $k$ D4-branes on each of the cycles. This gives
the $\prod_{i=1}^n U(k)_i$ gauge group and the bifundamentals. 
To reproduce the $v_i$ hypermultiplets, we wrap
$v_i$ D4-branes on a 2-cycle dual to $[a_i,a_{i+1}]$.
The configuration space of the D4-branes wrapping the
$S^2$-cycles gives the moduli space of the theory. In
particular, their configuration on $K3$ parametrizes
the hypermultiplet moduli space while their positions
on $T^3$ span part of the vector multiplet moduli space.

We can now see that the largest Higgs branch of this theory
is the instanton moduli space. To go to this Higgs branch, we
move all the D4-brane to the same location on $T^3$ (this corresponds
to moving to the origin of the Coulomb branch and turning
off masses of the hypermultiplet fields.).  We can then
move the $n \times k$ D4-branes wrapping on the $n$ $S^2$-cycles
off toward the imaginary direction in $z$. The
$n \times k$ D4-branes are then reconnected into $k$ D4-branes
wrapping on a cylinder $S^1 \times S^1$, where the first $S^1$
is on the fiber and the second $S^1$ is the real part of $z$.
By the T-duality on this $S^1 \times S^1$, 
these D4-branes becomes D2-branes
localized on $K3$. On the other hand, the $v_i$ D4-branes wrapping
on the dual 2-cycles become D6-brane wrapping on the entire $K3$.
Since the configuration of the D2-branes parametrizes the
hypermultiplet moduli space (the last sentence in the previous
paragraph) and the D2-branes on the D6-branes are
the same as instantons on the D6-branes, it is clear that
the largest Higgs branch of this theory is the instanton moduli space of
degree $k$ of rank $m = \sum_i v_i$ vector bundle. With some more
work, one can show that the vector bundle is exactly
$V = \oplus {\cal R}_i^{\otimes v_i}$.

\bigskip
\noindent
{\bf Acknowledgements}

We thank O. Aharony, M. Douglas,
K. Hori, P. Mayr, Y. Oz, N. Seiberg, M. Strassler
and S.-T. Yau for valuable discussions.
In addition we wish to thank Physics Department of Rutgers
University for the hospitality.

The work of H.O. is supported in part by NSF grant PHY-951497 and DOE
grant DE-AC03-76SF00098. The work of C.V. is supported in part by
NSF grant PHY-92-18167.

\listrefs
\end